\documentclass[twocolumn,iop,jpcm]{revtex4-2}
\usepackage[latin9]{inputenc}
\usepackage{amsmath}
\usepackage{graphicx}
\begin{document}
\title{Universal Anomaly of Dynamics at Phase Transition Points Induced by
Pancharatnam-Berry Phase}
\author{Jia-Yuan Zhang,$^{1}$ Xia Yin,$^{2}$ Ming-Yu liu,$^{1}$ Jize Zhao,$^{3,4}$
Yang Ding$^{2}$ and Jun Chang$^{1,4}$}
\email[Email address: ]{junchang@snnu.edu.cn}

\affiliation{$^{1}$College of Physics and Information Technology, Shaanxi Normal
University, Xi'an 710119, China~\\
 $^{2}$Center for High-Pressure Science and Technology Advanced Research,
Beijing 100094, China~\\
 $^{3}$School of Physical Science and Technology\&Key Laboratory
for Magnetism and Magnetic Materials of the MoE, Lanzhou University,
Lanzhou 730000, China~\\
 $^{4}$Lanzhou Center for Theoretical Physics and Key Laboratory
of Theoretical Physics of Gansu Province, Lanzhou University, Lanzhou
730000, China}
\begin{abstract}
Dynamical anomalies are
often observed near both the continuous and first-order phase transition
points. We propose that the universal anomalies could originate from
the geometric phase effects. A Pancharatnam-Berry phase is accumulated
continuously in quantum states with the variation of tuning parameters.
Phase transitions are supposed to induce an abrupt shift of the geometric
phase. In our multi-level quantum model, the quantum interference
induced by the geometric phase could prolong or shorten the relaxation
times of excited states at phase transition points, which agrees with
the experiments, models under sudden quenches and our semi-classical
model. Furthermore, we find that by setting a phase shift of $\text{\ensuremath{\pi}}$,
the excited state could be decoupled from the ground state by quantum
cancellation so that the relaxation time even could diverge to infinity.
Our work introduces the geometric phase to the study of conventional
phase transitions as well as quantum phase transition, and could substantially
extend the dephasing time of qubits for quantum computing. 
\end{abstract}
\date{\today }

\maketitle
\section{Introduction}
Phase transitions are of crucial importance
in physics since a variety of static and dynamic properties of systems are changed 
\cite{Stanley1999,Kadanoff1967static,RevModPhys.49.435-Theory-of-dynamic-critical-phenomena,Heyl2018,dziarmaga2010dynamics}.
In a long history, the study of phase transition focuses
on the static thermodynamic properties in equilibrium states. Recently,
it was shown that dynamical measurements could provide a direct insight into the investigation
of the complex transitions \cite{collet2003laser,pressacco2021subpicosecond,PhysRevB.105.075113,zhu2018unconventional,PhysRevLett.114.216403,PhysRevB.86.035128w4,PhysRevLett.116.107001w2,PhysRevLett.112.117801,PhysRevLett.106.207402w50,PhysRevLett.124.095703}.
Remarkably, the slowing-down dynamics near the phase transition point
have been observed in solids \cite{PhysRevLett.123.097601Slowing-Down-Phase-Transition,horie1987first,zhu2018unconventional,PhysRevLett.124.095703},
glasses \cite{PhysRevLett.72.1283Glass} and even microbial systems
\cite{veraart2012recovery}. In the symmetry-breaking phase transition,
the critical slowing down under perturbation could clearly be observed
in both experiments and theoretical models at critical points \cite{PhysRevLett.123.097601Slowing-Down-Phase-Transition,PhysRevLett.79.5154Temperature,PhysRevLett.114.037204Temperature-2,PhysRevA.97.013853SlowingDownByPhaseT,PhysRevE.88.032708-enhancement-3}.
The divergence of the relaxation time is attributed to the divergent
correlation length according to the renormalization group theory \cite{PhysRevLett.56.416Enhancement}.
However, near the first-order phase transition, the ultrafast relaxation time from the photoexcited
state to the equilibrium state also increases by orders of magnitude
in charge-ordered LaSrFeO \cite{zhu2018unconventional}. Similarly,
 slowing-down dynamics were observed near the first-order Mott transition
and structural phase transition \cite{PhysRevLett.124.095703,horie1987first}.
More surprisingly, in the superconducting and antiferromagnetic phase
transitions, the lifetimes of the decay are even shortened at
the critical point \cite{PhysRevLett.116.107001w2,PhysRevLett.106.207402w50}.
Furthermore, in some 1D short-range spin models under sudden quenches,
the fastest relaxations are unexpectedly found at the critical points,
in contrast to the critical slowing down \cite{dziarmaga2010dynamics,PhysRevB.103.214402,PhysRevLett.103.056403,PhysRevLett.102.130603}.
Therefore, the dynamical anomalies near phase
transition points are expected to be universal phenomena in a vast
number of systems. 

In this paper, we propose that the universal anomalies of dynamics near phase transition points could originate from the effect
of the geometric phase.
Date back to 1956, Pancharatnam proposed that the relative phase between two polarized light beams determines the intensity of
the interferogram \cite{Pancharatnam1956generalized}. Later in 1984,
Berry realized that besides the dynamical phase, the
quantum state acquires a geometric phase in the adiabatic and
cyclic evolution of the time-dependent Hamiltonian \cite{Berry1984quantal}.
The geometric phase is then generalized by loosening the constraint of
adiabaticity, cyclicity and unity \cite{PhysRevLett.58.1593,Samuel1988},
and applied in many fields ranging from high-energy physics \cite{PhysRevLett.55.927}
to condensed matter \cite{PhysRevLett.71.657}, statistics \cite{PhysRevLett.53.722,PhysRevLett.55.2887},
molecular \cite{PhysRevLett.113.263004,PhysRevLett.128.030401}, ultracold
atoms \cite{RevModPhys.83.1523}, optics \cite{PhysRevLett.57.933,PhysRevLett.57.937}
and quantum computation \cite{ekert2000geometric}.

The quantum criticality have been investigated based on the geometric phase of ground states in the XY Spin model \cite{PhysRevLett.96.077206, PhysRevLett.95.157203}, the ground state overlap in Dicke mode \cite{PhysRevE.74.031123}, the ground-state energy and its derivative in Rabi model \cite{PhysRevA.104.063703}. However, in many-body systems, it is difficult or impossible to experimentally measure the ground state energy and the functional dependency of the geometric phase on parameters. Theoretically, the system Hamiltonian often could not be identified for a long time as in cuprates, iron pnictides, manganites and so on. In this paper, we prove the dynamical anomalies at phase transition points are universal as the result of quantum coherence, without the need for the exact system Hamiltonian or the precise dependency of the geometric phases on control parameters. Our model only includes some relevant quantum states of systems and coupling interactions. The information of the energy gaps between the states and the coupling constants could be measured via experiments. Therefore, the multi-level model is not limited to a special system.
We suppose that Pancharatnam-Berry phases appear in quantum states
and change continuously with tunable parameters such as external
fields in the Hamiltonians. In particular, such phases abruptly
change at phase transition points. Using the dissipative Schr\"odinger
equation, we study the geometric phase effects on the dynamical evolution of a multi-level system
near the phase transition point based on the generalized spin-boson
model. Near the phase transition points,  the relative phase between 
the states belonging to neighboring phases induces quantum interference, which results in the universal
anomalies of the dynamics. The dynamical anomalies could be applied
to probe the phase transition in experiments. We also use a semi-classical
model to corroborate the geometric phase effects on the relaxation
of the excited states. The relaxation time
at phase transition points could become longer or shorter, which is
in agreement with the experiments. Furthermore, we show that the relaxation
time even could go to infinity by setting some peculiar coupling parameters.
Our work could contribute to the study of phase transition and the design
of qubits with long dephasing time.

\begin{figure}
\includegraphics[width=0.4\paperwidth]{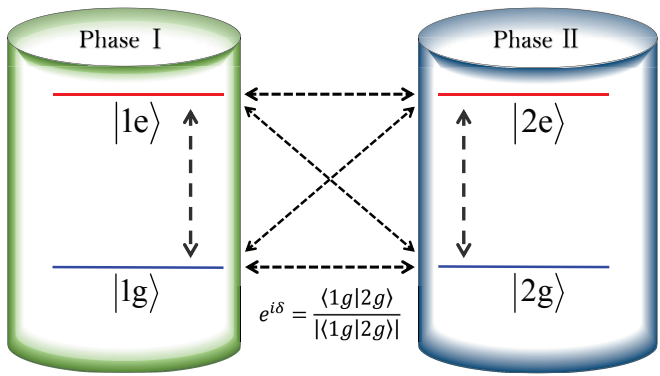}

\caption{The Schematic of the state coupling in Phase I , II and at the phase
transition points in a multi-level system. In Phase I and II, only
one excited state and one ground state are considered. At the phase
transition point, all the four states are coupled to each other due
to the phase fluctuation. We assume that the quantum state $\left|g_i\right\rangle $
in the $i$th phase acquires a continuous Pancharatnam-Berry phase
$\theta_{i}(\xi)$ with the change of the control parameters $\xi$.
The phase transition induces a geometric phase difference $\delta$
between the two ground states with $e^{i\delta}=e^{i(\theta_{2}-\theta_{1})}=\left\langle g_1\mid g_2\right\rangle /\left|\left\langle g_1\mid g_2\right\rangle \right|$
at $\xi_{c}$. Here, we have ignored the dynamical phase. \label{fig1} }
\end{figure}

\section{Quantum model}
We set up a quantum model to study the relaxation
process around phase transition points as shown in Fig. \ref{fig1}.
To elucidate the dynamical process, we introduce a system with the
Hamiltonian $H_{s}(\xi)$ under a phase transition from phase I to
phase II driven by the control parameter $\xi$, e.g. temperature,
pressure, magnetic field and interaction constants. For simplicity
and generality, we consider a multi-level system coupled to a bosonic
bath, i.e. a generalized spin-boson model. The model is first mapped
to another model with the electronic states coupled to a single
harmonic mode damped by an Ohmic bath \cite{PhysRevB.82.075124base,PhysRevLett.104.06740g_138,Ultrafast-X-ray}.
The Hamiltonian of the system is then written as

\begin{eqnarray}
H_{s}=\sum_{i}E_{i}c_{i}^{\dagger}c_{i}+\sum_{ij}V_{ij}(c_{i}^{\dagger}c_{j}+{\rm h.c.})\nonumber \\
+\sum_{i}\lambda_{i}c_{i}^{\dagger}c_{i}\left(a^{\dagger}+a\right)+\hbar\omega a^{\dagger}a,\label{h0}
\end{eqnarray}
where $E_{i}$ is the energy level and $n_{i}=c_{i}^{\dagger}c_{i}$
gives the occupation in the state $i$, $V_{ij}$ is the real coupling
(hybridization) constant between states $j$
and $i$, e.g. Heisenberg exchange interaction, spin-orbit coupling constant, atom-field coupling strength. $a^{\dagger}$ is the creation operator for the bosonic
mode with frequency $\omega$. We further define the electron-boson
self-energy $\varepsilon_{i}=\lambda_{i}^{2}/\hbar\omega$, and the
self-energy difference $\varepsilon_{ij}=\left(\lambda_{i}-\lambda_{j}\right)^{2}/\hbar\omega$
as well as the energy gap $\Delta_{ij}=\left(E_{i}-\varepsilon_{i}\right)-\left(E_{j}-\varepsilon_{j}\right)$
between two states. $E_{i}$, $V_{ij}$, $\omega$ and $\lambda_{i}$
are the functions of the tuning parameter $\xi$.

The Ohmic bath damping is introduced by a dissipative Schr\"odinger
equation, in which a dissipative operator $iD$ is added to the Hamiltonian
to describe the bath induced dissipation on the system \cite{PhysRevB.82.075124base}, 
\begin{equation}
i\hbar\frac{d\left|\psi\left(t\right)\right\rangle }{dt}=\left(H_{0}+iD\right)\left|\psi\left(t\right)\right\rangle ,
\end{equation}
where ${H_{0}}=e^{S}H_{s}e^{-S}$ is the Fr\"ohlich transformation of $H_{s}$ with 
$S=1/{\hbar \omega }\sum_{i}n_i\lambda_i\left(a^{\dagger}-a\right)$. The eigenvectors $|\psi_{in}\rangle$ of $H_{0}$ are selected as
the basis of the state $i$ with $n$ excited boson modes. We still
need the detailed time evolution formula for $P_{in}(t)$. On one hand, the coupling to the surroundings relaxes a state
with $n$ bosons to a $n-1$ boson state by the emission of bosons.
On the other hand, the probability of the $n$-boson state increases
due to the decay of the state with $n+1$ bosons. This gives a change
in the probability of the $n$-boson state 
\begin{eqnarray}
\frac{dP_{in}(t)}{dt}=-2n\bar{\Gamma}P_{in}(t)+2(n+1)\bar{\Gamma}P_{i,n+1},\label{rate}
\end{eqnarray}
where $P_{in}(t)=\left|\left\langle \psi_{in}\right|\left. \psi(t)\right\rangle \right|^{2}$,
$\bar{\Gamma}=\pi\bar{\rho}\bar{V}^{2}/\hbar$ is the environmental
relaxation constant, where $\bar{\rho}$ is the effective environmental
boson density of states and $\bar{V}$ is the interaction between
the local system and the environment.  The dissipative Schr\"odinger
equation effectively incorporates both the strong electron-boson coupling and
environment memory effects by introducing the bosonic mode in the
system, which were described previously \cite{PhysRevB.90.104305g31}. Details of the dissipative Schr\"odinger equation are provided in the Supplemental Materials (SM) \cite{SM}. 

\section{Dynamics at phase transition point}
The cascade decay in
a multi-level system could be effectively described by a two-level
system. One of them is the excited state and the other is the ground
state \cite{PhysRevB.82.075124base}. Therefore, in this paper, we
only consider the relaxation process in such a two-level system. We
assume that the Ohmic bath is the same in two different phases.
The ground states are represented by $\left|g_1\right\rangle $ and $\left|g_2\right\rangle $,
and the excited states $\left|e_1\right\rangle $ and $\left|e_2\right\rangle $
in phase I and II , respectively. At the phase transition point, we
assume the four states in phase I and II coexist due to phase fluctuation.
There are inter couplings between the states in phase I and II, such
as $V_{g_ie_j}$, $V_{g_ig_j}$ and $V_{e_ie_j}$ with $i\neq j$. Importantly,
we assume that the quantum state acquires a Pancharatnam-Berry phase
$\theta_{i}(\xi)$ with the change of the control parameters $\xi$
in the $i$th phase. The phase transition induces a geometric phase
difference between the two ground states with $e^{i\delta}=e^{i(\theta_{2}-\theta_{1})}=\left\langle g_1\mid g_2\right\rangle /\left|\left\langle g_1\mid g_2\right\rangle \right|$
at $\xi_{c}$. The total probability $P=P_e +P_g$ is normalized to 1 with $P_{e}=P_{e_1}+P_{e_2}$ for the exited states and $P_{g}=P{}_{g_1}+P_{g_2}$ for the  ground states. To
study the relaxation of excited states, we assume the initial state
is $\left|e_1\right\rangle $ or $\left|e_2\right\rangle $ adiabatically
excited from $\left|g_1\right\rangle $ or $\left|g_2\right\rangle $
in phase I or phase II, or the superposition of $\left|e_1\right\rangle $
and $\left|e_2\right\rangle $ at the phase transition point. Solving
the dissipative Schr\"odinger equation numerically, the evolution of
all the states as a function of time clearly reflects the dynamic
processes in phase I , II and at the phase transition point. In this
paper, we set $\hbar\omega$ of the single harmonic boson mode as
the energy unit, and $\tau=2\pi/\omega$ as the unit of time.

\begin{figure}
\includegraphics[width=0.4\paperwidth]{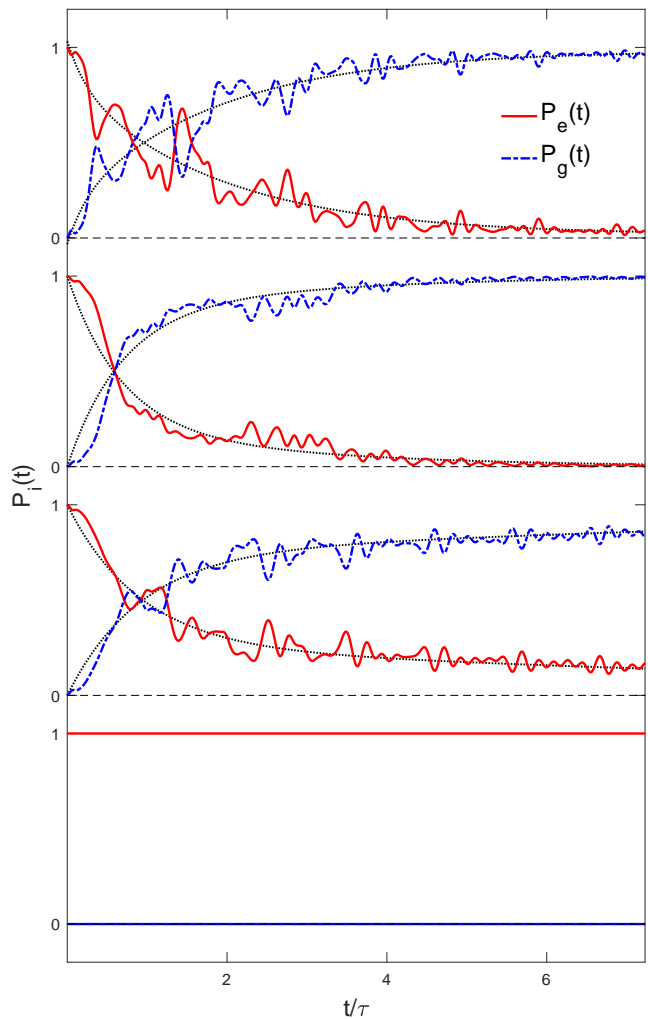}

\caption{The time evolution of state probabilities. $P_{e}$ and $P_{g}$ are
the probabilities of the excited states and ground states, respectively. The dotted curves are obtained by fitting to the exponential function $ y(t)=y_0+a_1*e^{-t/t_1}+a_2*e^{-t/t_2}$.
The starting state $\left|e\right\rangle $ is excited from $\left|g\right\rangle $
adiabatically. At the phase transition point, the starting state is
the hybridization of $\left|e_1\right\rangle $ and $\left|e_2\right\rangle $
with the same probability. $\hbar\omega=0.06$ eV of the single harmonic
boson mode is set as the unit of energy, and $\tau=2\pi/\omega\approx69$
femtosecond (fs) as the unit of time (more details can be found in the SM ). We take the environmental relaxation
time $(2\bar{\Gamma})^{-1}=50 fs\approx0.7\tau$, as in pervious
paper \cite{PhysRevLett.104.06740g_138}. (a) the probability evolution in
phase I or II, $\Delta_{g_1e_1}=\Delta_{g_2e_2}=5$, the coupling between
the ground state and exited state $V_{g_1e_1}=V_{g_2e_2}=1/2$. The electron-boson
coupling constants $\lambda_{g_1}=-\lambda_{g_2}=1/\sqrt{2}$ and $\lambda_{e_1}=-\lambda_{e_2}=\sqrt{5}$.
(b) at the phase transition point with $\delta=0$, the interstate coupling
$V_{g_1e_2}=V_{g_2e_1}=1/4$ and the hybridization $V_{g_1g_2}=V_{e_1e_2}=0.1$.
(c) at the phase transition point with $\delta=\pi$, other parameters
are the same as (b). (d) $\delta=\pi$ and $V_{g_1e_2}=V_{g_2e_1}=V_{g_1e_1}=V_{g_2e_2}=1/2$,
$\lambda_{g_1}=\lambda_{g_2}=1/\sqrt{2}$ and $\lambda_{e_1}=\lambda_{e_2}=\sqrt{5}$
and other parameters are the same as (b) and (c). \label{fig2}}
\end{figure}

To underscore the dynamical anomaly at the phase transition points,
firstly, we set the parameters $E_{i}$, $\lambda_{i}$ and $V_{ij}$
to ensure that the decay processes are the same in both phases. We
assume $E_{i}$, $V_{ij}$ and $\omega$ are independent of the control
parameter $\xi$, and only $\lambda_{i}$ may change with $\xi$. For example, the abrupt change of the $\lambda_{i}$ with respect to $\xi$ could be selected as the order parameter of phase transition. 
The energies of the ground states $E_{g_1}$ and $E_{g_2}$ are set to be zero
in both phases, and the energies of the excited states $E_{e_1}=E{}_{e_2}$.
The coupling constants between the ground state and excited state
are the same $V_{g_1e_1}=V_{g_2e_2}$. For the electron-boson coupling,
we take $\lambda_{g_1}=-\lambda_{g_2}$ and $\lambda_{e_1}=-\lambda_{e_2}$,
and hence $\varepsilon_{g_1}=\varepsilon_{g_2}$, $\varepsilon_{e_1}=\varepsilon_{e_2}$
and $\Delta_{g_1e_1}=\Delta_{g_2e_2}$. Therefore, the exited states in
the both phases decay in the same way, as shown in Fig. \ref{fig2}(a).
On the other hand, at the phase transition point, we assume that the
four states coexist due to the fluctuation. The intercoupling constants
between the ground states and exited states are supposed to be the
same $V_{g_1e_2}=V_{g_2e_1}$. Furthermore, there are fluctuations
within the ground and exited states with $V_{g_1g_2}=V_{e_1e_2}$. The
starting states are excited adiabatically from the ground states $\left|g_1\right\rangle $,
$\left|g_2\right\rangle $ or the mixture. We assume that there is
a Pancharatnam-Berry phase difference $\delta$ between the ground
state $\left|g_1\right\rangle $ and $\left|g_2\right\rangle $ of phase
I and II. It is expected that the relaxation time at the phase transition
point should be quite close to that in phase I and II. However, we
find that the relaxation strongly depends on the relative phase $\delta$
due to quantum interference effects. When there is no phase difference
or $\delta=0$, the relaxation time at the phase transition point
is even slightly shorter than that in phase I or II as shown in Fig.
\ref{fig2}(b), which is in agreement with the reduction of relaxation
time observed in the experiment at the critical point \cite{PhysRevLett.116.107001w2,PhysRevLett.106.207402w50}.
On the other hand, in Fig. \ref{fig2}(c) for $\delta=\pi$, the decay
time at the phase transition point could be much longer than that
in phase I or II, as the slowing down observed in many experiments
\cite{zhu2018unconventional,PhysRevLett.123.097601Slowing-Down-Phase-Transition,PhysRevLett.123.203602slowdown,PhysRevLett.110.257601Critical-slowing-down-light-1,PhysRevLett.114.037204Temperature-2}.

More surprisingly, when we set $\delta=\pi$, $V_{g_1e_2}=V_{g_2e_1}=V_{g_1e_1}=V_{g_2e_2}$,
$\lambda_{g_1}=\lambda_{g_2}$, $\lambda_{e_1}=\lambda_{e_2}$ and keep
the rest of the parameters the same as in Fig \ref{fig2}(c), the relaxation
time of the exited states at the transition point even could stretch
into infinity as shown in Fig. \ref{fig2}(d), which is similar to the
critical slowing down in the continuous phase transitions but it is
independent of the divergent correlation length. As a contrast, for
$\delta=0$, the relaxation time of the excited states is close to
that in phase I or II (not shown). Actually, it has been realized
in experiments since several decades ago that the superconducting
qubits composed by Josephson junctions with $\pi$ phase shifters
could be efficiently decoupled from environments and extend the phase
coherence time \cite{Josephson-junctions,ConclusionOfJosephsonJunction,Feofanov2010}.
To apprehend this puzzling result, we consider a four-level system
without a bath. We assume that $\left|g_1,g_2\right\rangle $ are the
wave vectors for the ground states and $\left|e_1,e_2\right\rangle $
the wave vectors for the exited states. The time-dependent Schr\"odinger
equation of the four states is written as 
\begin{equation}
i\hbar\frac{d}{dt}\left[\begin{array}{c}
\left|e_1\right\rangle \\
\left|e_2\right\rangle \\
\left|g_1\right\rangle \\
\left|g_2\right\rangle 
\end{array}\right]=\left[\begin{array}{cccc}
E_{e} & v & V & V\\
v & E_{e} & V & V\\
V & V & E_{g} & v\\
V & V & v & E_{g}
\end{array}\right]\left[\begin{array}{c}
\left|e_1\right\rangle \\
\left|e_2\right\rangle \\
\left|g_1\right\rangle \\
\left|g_2\right\rangle 
\end{array}\right],
\end{equation}
where $E_{e,g}$ are the energies of the four states, $v$ and $V$
are the state coupling constants. One has

\begin{equation}
i\hbar\frac{d\left|g_1\right\rangle }{dt}=E_{g}\left|g_1\right\rangle +v\left|g_2\right\rangle +V\left|e_1\right\rangle +V\left|e_2\right\rangle .\label{qcancell}
\end{equation}
When $\left|e_1\right\rangle $ has a $\pi$ phase difference with
respect to $\left|e_2\right\rangle $, i.e. $\left|e_1\right\rangle =e^{i\pi}\left|e_2\right\rangle $,
then the two last terms in Eq. (\ref{qcancell}) cancel each other
and the ground states are decoupled from the two exited states. Interestingly,
Fig. \ref{fig2}(d) indicates that even the environmental dissipation
is involved, the excited states and ground states still could be decoupled
from each other by canceling. Consequently, the quantum cancellation
induced by the $\pi$ phase shift is the key to extending the phase
coherence time in the experiments \cite{Josephson-junctions,ConclusionOfJosephsonJunction,Feofanov2010}. 

\section{Semiclassical relaxation model}
To qualitatively understand
the time relaxation in the quantum model, we appeal to a semiclassical
model. A general phenomenological model is proposed to study the time
evolution of the excited states near the phase transition points.
We assume that $\left|e_1,e_2\right\rangle $ are the wave vectors of
the excited states in the neighboring phase I and phase II. At the
phase transition point, two excited states coexist and are weakly
coupled to each other due to phase fluctuation. The quantum coherence
is set up between the two excited states. We study the time evolution
of the two exited states by mimicking the method in the Feynman's
phenomenological model of the Josephson junction \cite{SM}. 

\begin{equation}
i\hbar\frac{\partial\left|e_1\right\rangle }{\partial t}=(E_{e_1}-i\Gamma_{1})\left|e_1\right\rangle +(K-iK')\left|e_2\right\rangle ,
\end{equation}

\begin{equation}
i\hbar\frac{\partial\left|e_2\right\rangle }{\partial t}=(E_{e_2}-i\Gamma_{2})\left|e_2\right\rangle +(K-iK')\left|e_1\right\rangle ,
\end{equation}
where $E_{e_1}, E_{e_2}$ are the energies, $\Gamma_{1},\Gamma_{\ensuremath{2}}$
are the relaxation rates of the two exited states, respectively. While
$K$ is the state coupling constant and $K'$ is relaxation rate constant.
If $K$ and $K'$ are zero, then the two Schr\"odinger equations describe
the two excited states in the phase I and II, respectively. Near the
critical point, the coupling or fluctuation between the two states
may induce tunneling from one state to the other. Defining the total
excited quasiparticle density $P_{e}=P_{e_1}+P_{e_2}$ with $P_{e_1}=\left\langle e_1\mid e_1\right\rangle $
and $P_{e_2}=\left\langle e_2\mid e_2\right\rangle $, and the phase difference
$e^{i\delta}=\left\langle e_1\mid e_2\right\rangle /\left|\left\langle e_1\mid e_2\right\rangle \right|$,
then one has 
\begin{equation}
\hbar\frac{\partial P_{e}}{\partial t}=-2\Gamma'P_{e},
\end{equation}
with the effective relaxation rate 
\begin{equation}
\Gamma'=\frac{\Gamma_{1}P_{e_1}+\Gamma_{2}P_{e_2}}{P_{e}}+\alpha K'\cos\delta,\label{Gammap}
\end{equation}
where $\alpha=2\left|\left\langle e_1\mid e_2\right\rangle \right|/P_{e}$
and $0\leq\alpha\leq1$. Since $\max(\Gamma_{1},\Gamma_{2})>\left(\Gamma_{1}P_{e_1}+\Gamma_{2}P_{e_2}\right)/P_{e}>\min(\Gamma_{1},\Gamma_{2})$,
the relaxation rate should change smoothly from one phase to the other
if the quantum coherence in the last term of Eq. (\ref{Gammap}) is
ignored. However, the quantum coherence could strongly affect the
relaxation rate. For example, if we further assume $K'=\Gamma_{1}=\Gamma_{2}=\Gamma$,
then, the effective relaxation rate reads,

\begin{equation}
\Gamma'=\left(1+\alpha\cos\delta\right)\Gamma.\label{cGamma}
\end{equation}
Since $\cos\delta$ could be zero, positive or negative, one has $0\leq\Gamma'\leq2\Gamma$.
When $\delta=0$, $\Gamma'$ is larger than $\Gamma$. While $\Gamma'=0$
signifies the relaxation time $\tau'=1/\Gamma'$ approaching infinity,
similar to the critical slowing down.

\begin{figure}
\includegraphics[width=0.4\paperwidth]{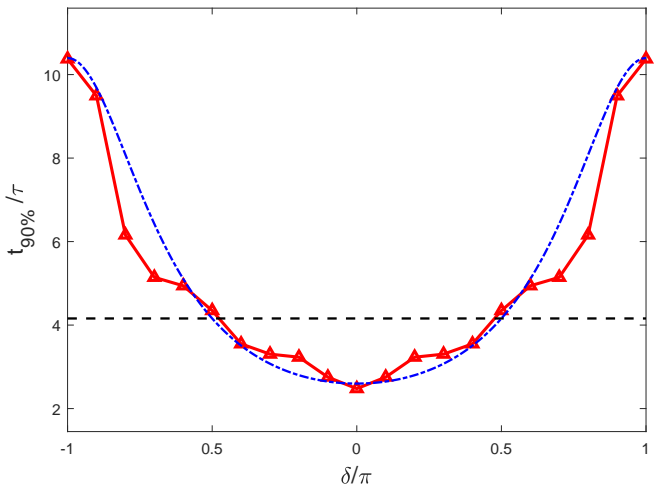}\\

\caption{The time $t_{90\%}$ for the ground state reaching its $90$\% population
as a function of the geometric phase difference $\delta$ at phase
transition points. The relaxation times at the critical point could
be longer or shorter than that in a single phase, in agreement with
the experiments. The model parameters are the same as Fig. \ref{fig2}(c).
Due to the fluctuations of the probability with time, the calculation results of the dissipative Schr\"odinger equation are fitted to exponential functions as shown in Fig. \ref{fig2}, and the triangles collected by solid red line denote $t_{90\%}$ obtained from the fitting curves. With the change of the relative
phase $\delta$, the decay times of the excited states vary roughly
from $4\tau$ to $11\tau$. The dashed blue curve is given by the
semiclassical model at phase transition point with $t_{90\%}=-\hbar\ln0.1/2\Gamma'$,
where $\Gamma'=\left(1+\alpha\cos\delta\right)\Gamma$ according to
Eq. (\ref{cGamma}) with $\alpha=0.6$. In the semiclassical model,
$\Gamma$ is calculated by Fermi's golden rule in Eq. (\ref{qgamma})
using the parameters in the quantum model. In the phase I or II, $t_{90\%}=-\hbar\ln0.1/2\Gamma\approx4.16\tau$
is a constant denoted by the horizontal dashed black line. \label{fig3}}
\end{figure}

In order to compare the quantum and semiclassical models, we study
the influence of the relative phase $\delta$ on the relaxation time
quantitatively in both models. We define $t_{90\%}$ as the time for
the ground state reaching $90$ percent of the total population. As
shown in Fig. \ref{fig3}, the dependence of $t_{90\%}$ on $\delta$
in the quantum model agrees well with that in the semiclassical
model. With $\delta$ varying from 0 to $\pi$, the decay rate decreases
gradually and the value of $t_{90\%}$ is strongly enhanced. For instance,
when there is no phase difference or $\delta$=0, $t_{90\%}$ at the
phase transition point is even slightly shorter than that in phase
I or II. On the other hand, for $\delta=\pi$, $t_{90\%}$ at the
phase transition point is much longer than that in phase I or II.
In the semiclassical model, using the parameters in the quantum model,
$\Gamma$ of the excited states is calculated by Fermi's golden rule
\cite{PhysRevB.82.075124base}

\begin{equation}
\Gamma=\frac{\pi F_{n}V_{ge}^{2}}{\hbar\omega}\label{qgamma}
\end{equation}
where $V_{ge}$ is the coupling strength between the ground state
and excited state and $F_{n}=e^{-g}g^{n}/n!$ is the Franck-Condon
factor with $n\approx\Delta_{ge}/\hbar\omega$ with energy gap $\Delta_{ge}=\left(E_{e}-\lambda_{e}^{2}/\hbar\omega\right)-\left(E_{g}-\lambda_{g}^{2}/\hbar\omega\right)$,
and $g=\varepsilon_{ge}/\hbar\omega$ is the Huang-Rhys factor with
the electron-phonon selfenergy difference $\varepsilon_{ge}=\left(\lambda_{e}-\lambda_{g}\right)^{2}/\hbar\omega$.
At phase transition point, the decay time $t_{90\%}=-\hbar\ln0.1/2\Gamma'$
with $\Gamma'=\left(1+\alpha\cos\delta\right)\Gamma$. In the phase
I or II, $t_{90\%}=-\hbar\ln0.1/2\Gamma$ is a constant around $4.16$$\tau$.

\section{Discussion and Conclusions}
The phase difference between the ground state and the excited state is not always exactly the same. For example, in the XX spin model, the phase difference between the ground state and the excited state changes from $0$ to $\pi$ near the critical point \cite{PhysRevLett.95.157203}. Interestingly, the anomaly of the relaxation time dominantly originates from the relative phase difference between the two excited states belonging to the phase I and II, respectively, which could be roughly understood from our semiclassical model and the Eq. (\ref{qcancell}) of the 4-level model without dissipation. Furthermore, within the single phase I or II, the phase difference between the ground state and the excited state has limit effects on the relaxation time. The relative phase between different states could be tuned by external fields. 

To conclude, we have proposed that the conventional
phase transition and quantum phase transition could be featured by
the Pancharatnam-Berry phase factor. We assumed that with the change
of the control parameter in the Hamiltonian, the quantum state accumulates
a geometric phase and there is abrupt shift of the phase at
phase transition points. Applying the dissipative Schr\"odinger equation,
we studied the dynamical evolution of the generalized spin-boson model.
At the phase transition point, the geometric phase difference between
the states belonging to neighboring phases results in the universal anomalies
of dynamics via quantum interference. Since the geometric phase strongly
affect the dynamical relaxation near phase transition points, experimental
measurements of the dynamical anomalies could be applied to probe
the phase transition. The effects of the geometric phase on the relaxation
times in the quantum model qualitatively agrees with our semiclassical
model. The geometric phase can increase or decrease the relaxation
time at phase transition points, which coincides with the experiments
and existing models. Furthermore, by adjusting some parameters and
setting a $\text{\ensuremath{\pi}}$ phase shift, we found that the
relaxation time of the excited states even could be divergent, which
agrees well with experiments of the superconducting $\pi$-junction.
Our work presented theoretical evidences for studying the phase transition
by introducing the geometric phase, which is benefited for the design
of qubits with a long dephasing time for quantum computation and communications.

\textit{Acknowledgments}.$-$ This work is supported by the National
Natural Science Foundation of China (Grants No.12274187, No.11874188, No.12047501, No.11874075), Science Challenge Project No. U1930401, and National Key Research and Development Program of China 2018YFA0305703.

\end{document}